\documentclass{PoS}

\usepackage{lineno}
\usepackage{upgreek}
\usepackage{url}
\usepackage{caption}
\newcommand{\change}[1]{{#1}}

\title{TARGET: toward a solution for the readout electronics of the Cherenkov Telescope Array}

\ShortTitle{The TARGET readout ASIC for CTA}

\author{\speaker{L. Tibaldo}$^a$, J.~A. Vandenbroucke$^b$, A.~M. Albert$^a$, S.~Funk$^{ca}$, T.~Kawashima$^d$, M.~Kraus$^{c}$, A.~Okumura$^{de}$, L. Sapozhnikov$^a$, H.~Tajima$^d$, G.~S. Varner$^f$, T.~Wu$^b$, A.~Zink$^c$, for the CTA consortium\footnote{Full consortium author list at http://cta-observatory.org}\\
E-mail: \email{ltibaldo@slac.stanford.edu}

{\footnotesize
        $^a$ Kavli Institute for Particle Astrophysics and Cosmology, SLAC National Accelerator Laboratory, Stanford University, Stanford, CA 94305, USA\\
	 $^b$ Department of Physics and Wisconsin IceCube Particle Astrophysics Center, University of Wisconsin, Madison, WI 53706, USA\\
	$^c$ Erlangen Centre for Astroparticle Physics, Friedrich-Alexander-Universit\"at Erlangen-N\"urnberg, Erwin-Rommel-Str. 1, D-91058 Erlangen, Germany\\
	$^d$ Solar-Terrestrial Environment Laboratory, Nagoya University, Furo-cho, Chikusa-ku, Nagoya, Aichi 464-8601, Japan\\
	$^e$ \change{Max-Planck-Institut f\"ur Kernphysik, P.O. Box 103980, D 69029 Heidelberg, Germany}\\
	$^f$ Department of Physics and Astronomy, University of Hawaii, 2505 Correa Road, Honolulu, HI 96822, USA\\
	}}
        
\abstract{TARGET is an application specific integrated circuit (ASIC) designed to read out signals recorded by the photosensors in cameras of very-high-energy gamma-ray telescopes exploiting the imaging of Cherenkov radiation from atmospheric showers. TARGET capabilities include sampling at a high rate (typically 1 GSample/s), digitization, and triggering on the sum of four adjacent pixels. The small size, large number of channels read out per ASIC (16), low cost per channel, and deep buffer for trigger latency (${\sim}16$~$\upmu$s at 1 GSample/s) make TARGET ideally suited for the readout in systems with a large number of telescopes instrumented with compact photosensors like multi-anode or silicon photomultipliers combined with dual-mirror optics. The possible advantages of such systems are better sensitivity, a larger field of view, and improved angular resolution. The two latest generations of TARGET ASICs, TARGET 5 and TARGET 7, are soon to be used for the first time in two prototypes of small-sized and medium-sized dual-mirror telescopes proposed in the framework of the Cherenkov Telescope Array (CTA) project. In this contribution we report on the performance of the TARGET ASICs and discuss future developments.}

\FullConference{The 34th International Cosmic Ray Conference,\\
		30 July- 6 August, 2015\\
		The Hague, The Netherlands}

\begin{document}

\section{TARGET: an ASIC for imaging atmospheric Cherenkov telescopes}

One the most effective techniques to detect gamma~rays at very high energies is the imaging of Cherenkov light from atmospheric showers. This field is soon to be revolutionized by the advent of the Cherenkov Telescope Array (CTA), which is meant to improve the sensitivity by about an order of magnitude, to extend the energy range, from a few tens of GeV to above 100 TeV, with enhanced angular and energy resolutions over existing imaging atmospheric Cherenkov \change{telescope} (IACT) systems  \cite{cta-paper,cta-paper2}.

The large number of telescopes required to reach the performance aimed at for CTA calls for innovative ways to reduce their cost. One of the proposed solutions is using dual-mirror optics along with a compact camera design enabled by modern photosensors like multi- anode or silicon photomultipliers. This helps to bring down the costs by reducing the area that needs to be instrumented with photosensors to achieve the desired field of view of $\sim$10$^\circ$. Additionally, this enables the realization of finely-pixelated cameras which provide possible advantages such as high angular resolution resulting in excellent source sensitivity \cite{vassiliev}. 

TARGET is an application-specific integrated circuit (ASIC) conceived to provide an affordable and reliable solution for the processing of the photodetector signals for these compact IACT cameras \cite{T1-paper,T1-ICRC}. High-channel density and integration of different key functionalities in TARGET are key to lowering the costs for the camera electronics. 

Each TARGET ASIC reads~out 16 channels (i.e., 16 photodetector pixels) in parallel.  The signal is sampled at a high and configurable rate, typically 1 GSa/s, and the samples are stored in a deep (16,384 samples per channnel) analog buffer. The samples are then digitized on demand. TARGET also provides a trigger signal based on the analog sum of the signal from four adjacent photodetector pixels, which can be used by higher-level logic to decide which signals to \change{be digitized} and read out.  One or several ASICs can be controlled and read out by a single companion field-programmable gate array (FPGA).

TARGET enabled us to develop 64-channel front-end electronics modules for 
compact IACT cameras using only four ASICs, one FPGA and two printed circuit boards. The low number of components is critical to reduce costs and increase reliability. The fifth and seventh generations of TARGET ASICs, TARGET 5 and TARGET 7, will be used for the first time in IACTs that constitute prototypes of two telescopes proposed for CTA: the Gamma-ray Cherenkov Telescope (GCT) \cite{icrc-gctcam} and the Schwarzschild--Couder Telescope (SCT) \cite{icrc-sctcam}.  

The first part of this paper focuses on TARGET 5 and TARGET 7, their performance, and the lessons learned from their testing and  characterization. \change{TARGET~7 was designed to improve the performance over TARGET~5 in terms of both signal sampling/digitization and  of trigger generation. While the first goal was achieved, the trigger performance of TARGET~7 has limitations intrinsic to the design of the ASIC. Thus, we designed a new generation of TARGET ASICs, TARGET C, with a separate companion ASIC for triggering, CCTV, that are described in the second part of the paper}. A small \change{number} of TARGET~C and CCTV ASICs \change{have} been recently manufactured and will be fully characterized in the coming months.

\section{TARGET 5 and TARGET 7}

\begin{figure}
\includegraphics[width=1\textwidth]{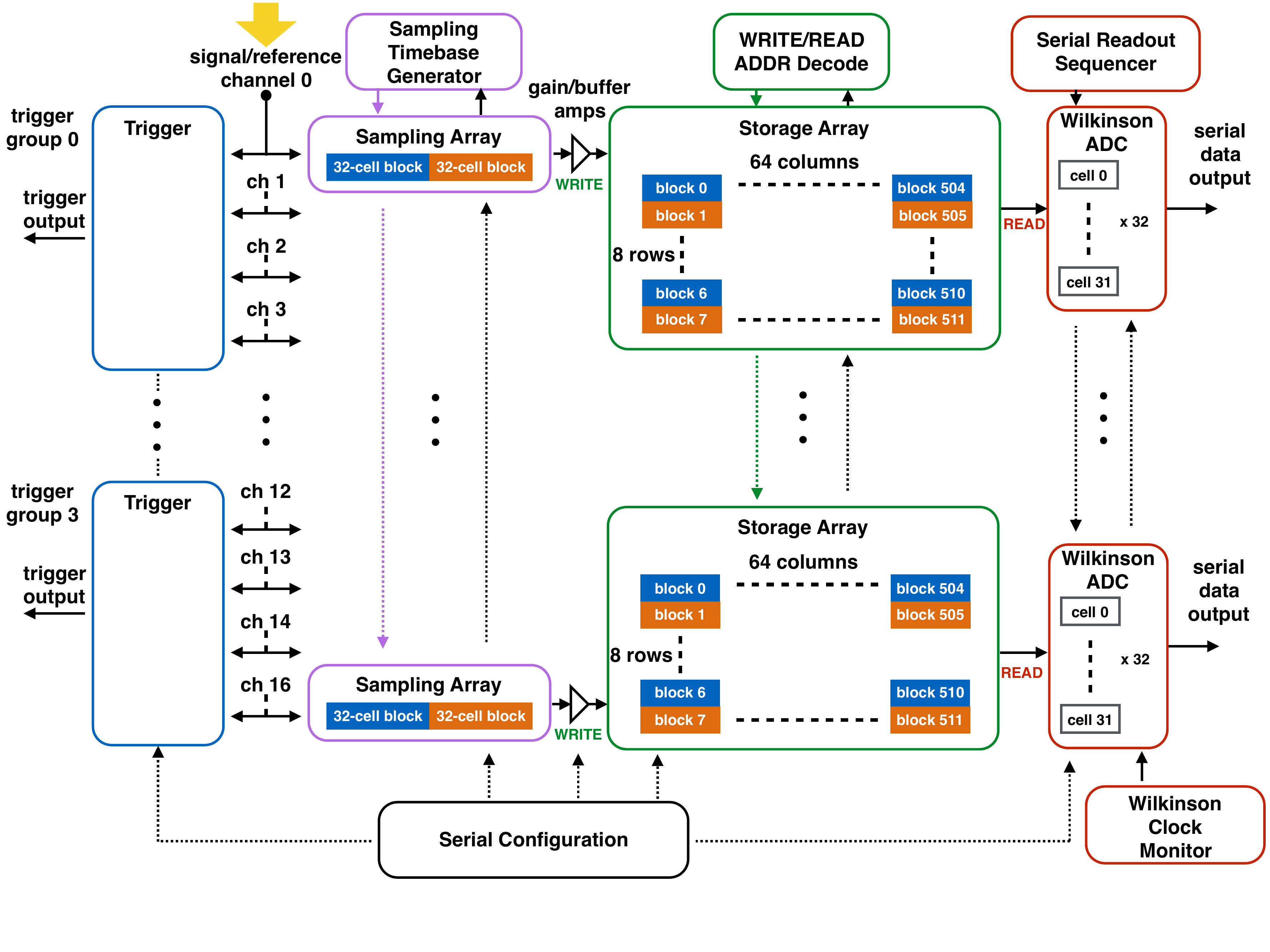}
\caption{Functional block diagram of the TARGET~5/TARGET~7 ASIC, with key
  components shown.  Sixteen channels are processed both for trigger
  formation, as well as analog sampling.  A timebase generator
  controls the sampling of signals into dual groups of 32 sampling
  cells.  Ping-pong operation transfers samples from one group of 32
  to storage while the other group is sampling, with the roles
  reversed in the next half sampling cycle.  These groups of 32
  storage cells can be randomly accessed on demand for readout by an onboard
  Wilkinson ADC on 32 samples in parallel for all 16 channels.
  Individual converted samples may then be selected and serially
  transmitted off-ASIC on all 16 channels concurrently.  In parallel the analog signals from groups of 4 adjacent channels are used to form triggers that are provided to off-ASIC high-level logic to generate the readout requests. Configuration
  of operating parameters are
  programmed through a serial-parallel interface.
}\label{T5blockdiagram}
\end{figure}

The input to \change{a TARGET ASIC} consists for each channel of both a signal line and a reference tied to a pedestal voltage to provide a DC offset. The ASIC provides independent paths for signal processing (sampling, digitization, and readout), the \emph{data path}, and triggering, \emph{the trigger path}. A schematic view of the ASIC functional blocks is shown in Fig.~\ref{T5blockdiagram}\change{.}

We characterized the TARGET ASICs using dedicated evaluation boards as well as prototypes of front-end electronics modules designed for GCT and SCT cameras \cite{icrc-gctcam,icrc-sctcam}. The characterization included \change{tuning the configurable parameters of the ASICs} to determine optimal \change{settings} for use in these IACT cameras.

\subsection{Data path}

The signal is first sampled by a 64-cell switched-capacitor array (consisting of two blocks operated in ping-pong fashion) and then buffered to a storage array of 16,384 switched-capacitor cells.  Samples from one group of 32 in the primary buffer are transferred to storage while the other group is sampling, with the roles reversed in the next half sampling cycle. This limits the input impedance and enables a large bandwidth while ensuring continuous sampling and large trigger latency ($\sim16$~$\upmu$s at 1 GSa/s). The timebase generator is an array of base delay elements with adjustable time step in the range from 1~ns to to ${\sim}2.5$~ns. For TARGET 5 the signals that govern the timebase generator are provided to the ASIC from an external source. In TARGET 7 they are generated internally in the ASIC to achieve higher sampling precision, hence reduced noise for the measurement of transient waveforms.  In both versions the sampling can be stabilized against temperature variation through control loops. 

\begin{figure}
\begin{tabular}{cc}
\includegraphics[width=.48\textwidth]{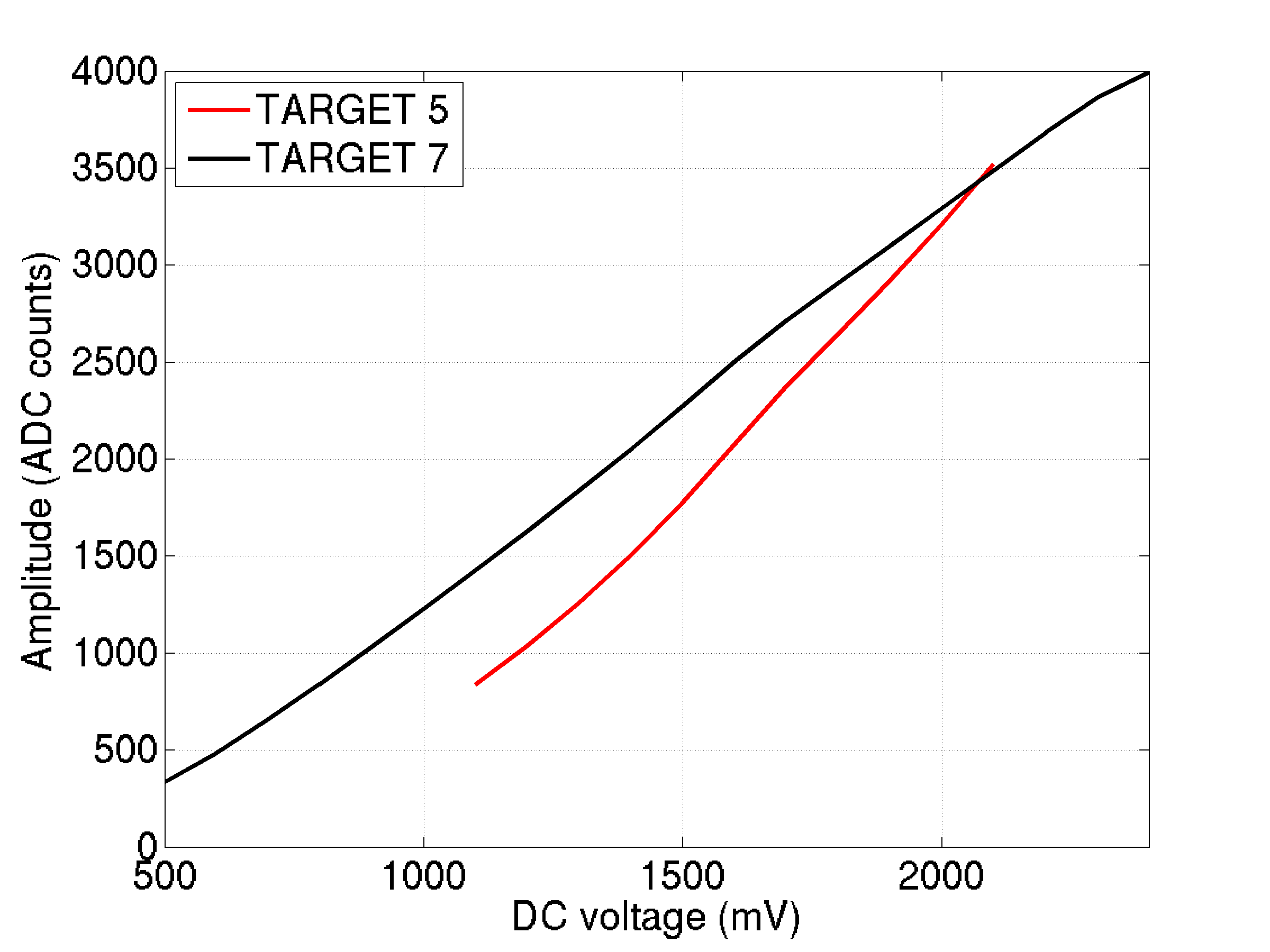}&
\includegraphics[width=.48\textwidth]{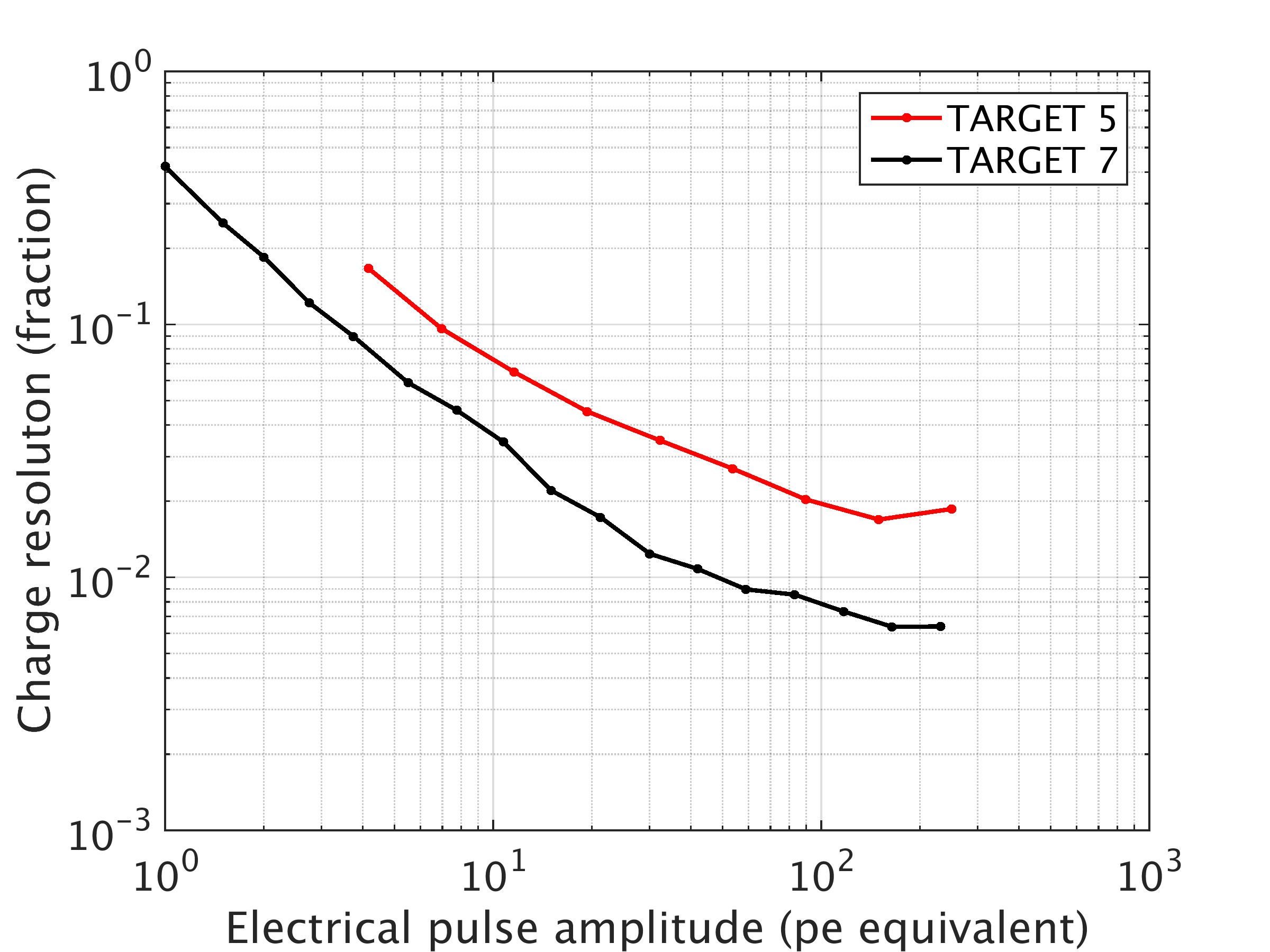}
\end{tabular}
\caption{Summary of the performance of the data path in TARGET~5 and TARGET~7: transfer function (left) and charge resolution (right). Note that the charge resolution includes the contribution from the sampling/digitization electronics  only, while additional contributions are expected from photodetectors and pre-amplifiers. The curves are shown up to saturation of the digitizer dynamic range.  Good performance can be continued beyond this by fitting saturated pulses.}
\label{fig-datapath}
\end{figure}

Samples stored in the secondary buffer are randomly accessible in blocks of 32 to perform digitization and \change{readout} on demand. Once selected, the 32 stored voltages are digitized through a highly configurable Wilkinson analog-to-digital converter (ADC). A
Wilkinson ramp generator block generates and broadcasts a ramp to all channels. At a second time a 12-bit ripple counter (with
adjustable speed) is started for each channel.  When the voltage ramp crosses the comparator
threshold for a given sample, the counter stops and the count then
represents the time (ADC code) corresponding to the voltage held in
the storage cell. Address decoding
and sequencing is performed inside a serial readout sequencer
block. Digitized samples are selected (again randomly accessible) and
then \change{serially} transferred on all 16 channels in parallel.
In TARGET 5 the clock for the Wilkinson counter is generated internally in the ASIC with an independent oscillator for each channel and an additional counter for monitoring and stabilizing the clock rate against temperature variation. 
In TARGET 7 an external clock signal along with an improved biasing scheme of the comparator in the Wilkinson ADC is used to improve the dynamic range and linearity of the transfer function (see later Fig.~\ref{fig-datapath}~left). However, this is at the price of slower conversion rate: a ``Done'' bit was added to compensate for the slower external digitization clock by enabling stopping the digitization and starting the data transfer as soon as all counters have reached their respective sample amplitudes.

The TARGET 5 dynamic range is 1.1~V and the TARGET 7 dynamic range is 1.9~V (Figure~\ref{fig-datapath}~left). The integral non linearity (maximum difference between the real transfer function, as shown in Figure~\ref{fig-datapath}~left, and the best-fit straight line) decreased from 75~mV in TARGET 5 to 40~mV in TARGET 7. With either ASIC, the nonlinearity is calibrated out using lookup tables that can be implemented in software or FPGA firmware.

We evaluated the contribution of TARGET electronic noise to the charge resolution using electrical pulses of known amplitude and shape from a function generator to simulate the signal from photodetectors after pre-amplification. It is assumed based on the planned photosensors and pre-amplifiers that the pulse amplitude arriving at TARGET is 4 mV per photoelectron (p.e.).  The charge resolution is shown in Figure~\ref{fig-datapath}~right.  The curves are measured up to saturation of the dynamic range.  Good performance can be continued beyond this by fitting saturated pulses.  We emphasize that this resolution only includes electronics contributions (dominated by noise but also including timing jitter and residual mis-calibration).  Additional contributions from photodetector noise, photodetector cross-talk, and Poisson noise will degrade the resolution.

\subsection{Trigger path}

The trigger path features an inverting summing amplifier that sums the analog signal in four channels composing a single trigger group.  A subsequent comparator applies an adjustable threshold to the analog sum.  The comparator output is routed to a one-shot to provide a trigger output signal with adjustable width. We characterized the trigger performance using function-generator pulses to measure the trigger efficiency as a function of input pulse amplitude and thereby determine the trigger threshold and noise.

\begin{figure}
\begin{tabular}{cc}
\includegraphics[width=.48\textwidth]{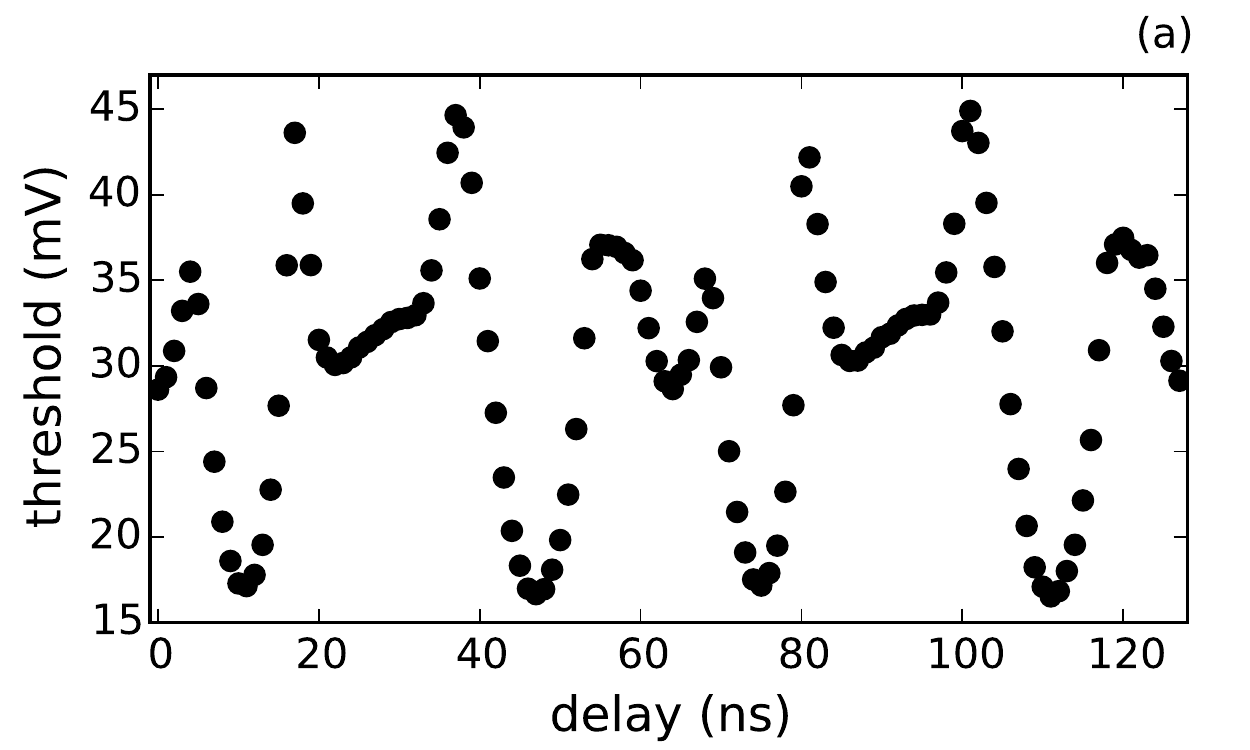}&
\includegraphics[width=.48\textwidth]{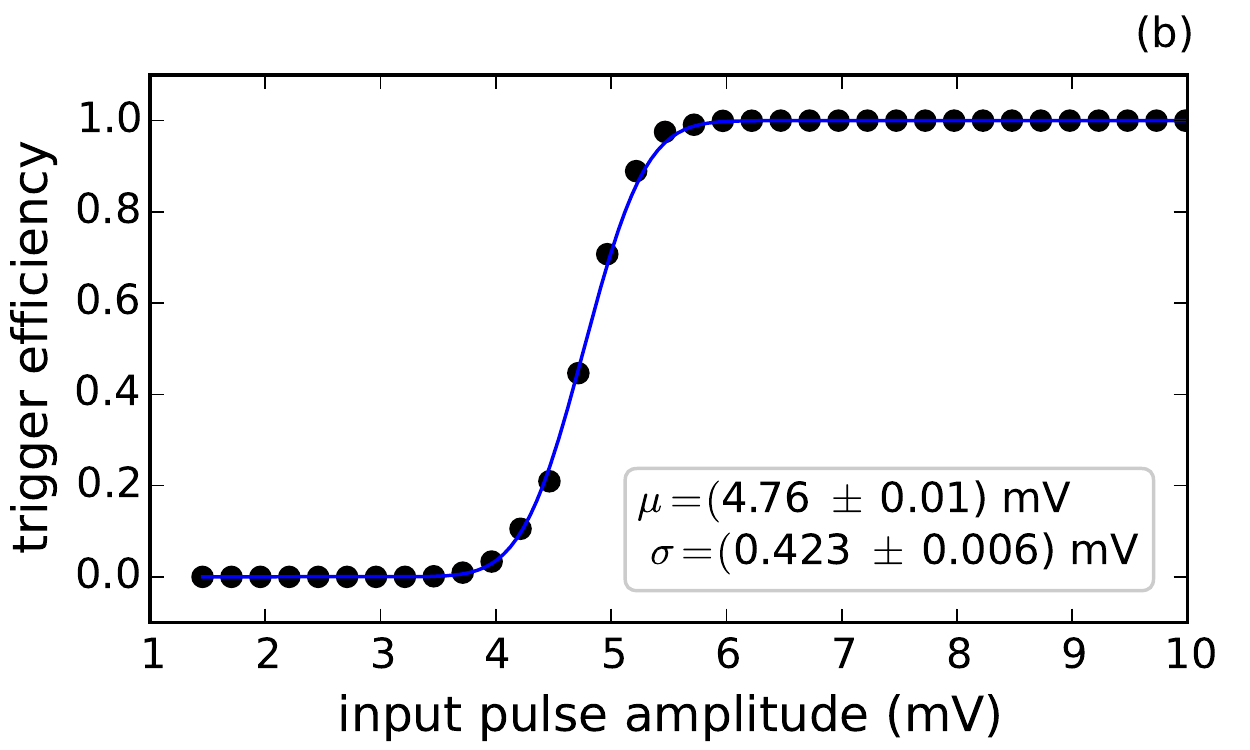}
\end{tabular}
\caption{TARGET 5 trigger performance. (a) Trigger threshold as a function of delay between \change{input pulse signal and sampling clock}, over two periods of the 64-ns sampling clock. (b) Demonstration of the low trigger threshold achievable with sampling disabled. The points show the trigger efficiency as a function of input pulse amplitude and the curve shows the best-fit Gaussian error function with mean (threshold) $\mu$ and standard deviation (noise) $\sigma$.}
\label{fig-trigger}
\end{figure}

For TARGET~5 during normal operations (i.e., with sampling ongoing on the ASIC) the minimum stable trigger threshold is 20~mV, and the typical trigger noise is ${\sim}5$~mV. If the injected pulse is synchronized with the \change{signals that drive the sampling timebase generator and the delay between the two is varied}, there are large periodic variations in trigger threshold   of 30~mV \change{with the 64-ns period of the sampling clock} (Figure~\ref{fig-trigger}a). Therefore, we concluded that noise related to sampling is the main source of the trigger noise observed in normal operations. When sampling is disabled\change{,} TARGET 5 has a tunable trigger threshold with a minimum of 4.5~mV (1.2~p.e.) and typical noise of $\sim0.5$~mV (0.13~p.e.) (Figure~\ref{fig-trigger}b).

To address noise picked up at the summing amplifier and comparator stages, in TARGET~7 the insulation of the trigger sector on the ASIC was improved, and each channel was equipped with two additional inverting amplification stages for the trigger path only before reaching the summing amplifier to boost the signal against pickup noise. Furthermore, in TARGET 7 the output trigger signal was improved to be a low-voltage differential signal (LVDS).

This design of TARGET-7 resulted in an inferior trigger performance, with minimum stable trigger threshold of 50~mV and typical trigger noise of ${\sim}15$~mV during normal operations. For signals synchronized with \change{the sampling clock there are variations in the trigger threshold of 50~mV as a function of relative phase between the two}. The larger effect in TARGET 7 despite the additional pre-amplification and improved insulation testifies that coupling between the sampling timebase generator and the trigger sector is likely to occur at a fundamental level through the silicon substrate of the ASIC. Although TARGET 5 and TARGET 7 perform sufficiently well for use in the prototype GCT and SCT cameras, further improvements have been implemented in a newest generation of ASIC.

\section{TARGET C and CCTV}

Due to the intrinsic coupling between trigger and sampling, we decided to separate the two functionalities in distinct ASICs. The new version of the ASIC performing sampling, digitization, and \change{readout}, named TARGET C, is similar to TARGET 7 with some modifications: a synchronous counter to reduce the digitization time; ability to tune individual delay steps in the timebase generator; and the reference point for the digitizer ramp was tied to ground for better noise control.

The triggering functionalities were moved to a companion ASIC named CCTV (Cherenkov Companion Trigger Variant). CCTV is a 16-channel device that features the trigger circuit from TARGET 5, except the trigger output is LVDS as in TARGET 7. CCTV also features an internal preamplifier that can be used to boost silicon photomultiplier signals and shape them with a full width at half maximum of $\sim$10~ns and amplitude of 4--8~mV/p.e., as achieved through external preamplifiers for TARGET 5 and TARGET 7.  The pre-amplified signal and a DC offset from CCTV can be used as input for any version of the TARGET ASIC from 5 on.  Alternatively, the CCTV preamplifier can be bypassed for backward-compatibility or for use with an external preamplifier.

A small batch of TARGET C and CCTV ASICs was recently manufactured  and their testing and characterization is currently ongoing. The performance of TARGET~C is expected to be very similar to TARGET 7 with faster digitization, while CCTV should achieve the trigger performance of TARGET 5 with sampling disabled.

\section{Summary and outlook}

\begin{table}
\small
\begin{center}
     \caption{Summary of the design characteristics and performance for TARGET 5, TARGET 7, and TARGET C + CCTV.  In all systems the nominal single-photoelectron amplitude is 4~mV.
}
     \label{sumtable}
     \begin{tabular}{lccc}
	\hline
			& \textbf{TARGET 5}	& \textbf{TARGET 7}	& \textbf{TARGET C + CCTV$^a$} \\
	\hline
	Characteristics\\
	\hline
	Number of channels 	& 16 	& 16 	& 16\\
	Sampling frequency (GSa/s) 	& $0.4-1$ 	& $0.5-1$ 	&  $0.5-1$\\
	Size of storage array (cells/channel) 	& 16,384 	& 16,384 	& 16,384\\
	Digitization clock frequency (MHz) & ${\sim}700$ & 208 & 500\\
	Samples digitized simultaneously	& $32 \times 16$ & $32 \times 16$ & $32 \times 16$ \\
	Trigger (sum of 4 channels) & integrated & integrated & companion\\
	\hline
	Performance\\
	\hline
	Dynamic range 	(V) & 1.1  & 1.9  & $\gtrsim1.9$ \\
	Integrated non linearity (mV) & 75 & 40 & $\lesssim 40$ \\
	charge linearity range (p.e.) & $4-300$ & $1-\gtrsim300$ & $1-\gtrsim 300$ \\
	charge resolution at 10 p.e. &  8\% & 4\% & 4\%$$ \\
	charge resolution at $>100$ p.e. & 2\% & $\lesssim 0.8\%$ & $\lesssim 0.8\%$\\
	Minimum trigger threshold (mV)$^b$ & 20 & 50 & 4.5$$\\
	Trigger noise (mV)$^b$ & 4 & 15 & 0.5$$\\
	\hline
     \end{tabular}
	\end{center}
	          {$^a$ The performance of TARGET C + CCTV is currently under evaluation: the numbers provided are expected based on studies of TARGET 7 and TARGET 5 with sampling disabled.\\
      $^b$ The trigger performance refers to normal operations, that for TARGET~5 and TARGET~7 includes signal sampling ongoing on the same ASIC.}
     \end{table}
     
Table~\ref{sumtable} summarizes the characteristics and performance of the latest generations of TARGET ASICs.
TARGET 5 and TARGET 7 are two important milestones for the readout systems planned for some CTA telescopes, and are being used in two prototypes that are in the construction/commissioning phase \cite{icrc-gctcam,icrc-sctcam}.

To overcome some limitations of the trigger performance, we have designed and produced two ASICs benefitting from experience with earlier generations, TARGET C for signal sampling, digitization and readout, and CCTV for providing trigger signals, which are currently ongoing characterization in the laboratory.  

\acknowledgments

We gratefully acknowledge support from the agencies and organizations listed under Funding Agencies at this website: \url{http://www.cta-observatory.org/}.


\begin{thebibliography}{99}

\bibitem{cta-paper} 
     M. Actis, G. Agnetta, F. Aharonian, et~al.
     \emph{Design concepts for the Cherenkov Telescope Array CTA: an advanced facility for ground-based high-energy gamma-ray astronomy},
     \emph{ExA} {\bf 32} (2011) 193
     [{\tt arXiv:1008.3703}].
     

\bibitem{cta-paper2}
      B.~S. Acharya, M. Actis, M.  {Aghajani}, et~al.
     \emph{Introducing the CTA concept},
     \emph{APh} {\bf 43} (2013) 3.
     
     
\bibitem{vassiliev}
	V.~.V Vassiliev, S. Fegan, P Brousseau
	\emph{Wide field aplanatic two-mirror telescopes for ground-based {$\gamma$}-ray astronomy},
	 \emph{APh} {\bf 28} (2007) 10

\bibitem{T1-paper} 
     K. Bechtol, S. Funk, A. Okumura, et~al.
     \emph{TARGET: A multi-channel digitizer chip for very-high-energy gamma-ray telescopes},
     \emph{APh} {\bf 36} (2012) 156
     [{\tt arXiv:1105.1832}].
     
\bibitem{T1-ICRC} 
     J.~A. Vandenbroucke, K. Bechtol, S. Funk, et~al.
     \emph{Development of an ASIC for Dual Mirror Telescopes of the Cherenkov Telescope Array},
     \emph{ICRC} (2011)
     [{\tt arXiv:1110.4692}].

\bibitem{icrc-gctcam}
	A. De~Franco, R. White, D. Allan, et.~al.
	\emph{The first GCT camera for the Cherenkov Telescope Array},
	\emph{PoS(ICRC2015)} 1015 (2015).
	
\bibitem{icrc-sctcam}
	A. N. Otte, J. Biteau, H. Dickinson, et.~al.
	\emph{Development of a SiPM Camera for a Schwarzschild-Couder Cherenkov Telescope for the Cherenkov Telescope Array},
	\emph{PoS(ICRC2015)} 1023 (2015).	
	 	
\end{thebibliography}
\end{document}